\documentclass[11pt,twoside]{article}  
\usepackage{apn3conf}

\begin{document}   

%
%
%
%

\title{Imaging and Spectroscopy of the Planetary Nebula NGC\,6778}


%
%
%

    \author{Vicente Maestro,
            Mart\'{\i}n A.\ Guerrero\altaffilmark{1},
            Luis F.\ Miranda}
    \affil{Instituto de Astrof\'{\i}sica de Andaluc\'{\i}a,
    Apdo.\ Correos 3004,
    E-18080, Granada, Spain}

    \altaffiltext{1}{Also at Department of Astronomy, University of Illinois}
    \altaffiltext{2}{Visiting Astronomer, Cerro Tololo Inter-American Observatory,
     National Optical Astronomy Observatory, operated by the Association of 
     Universities for Research in Astronomy, Inc., under a cooperative agreement 
     with the National Science Foundation.}

%
%

\contact{Vicente Maestro}
\email{vicente@iaa.es}

%
%
%
%
%

\paindex{Maestro, V.}
\aindex{Guerrero, M. A.}     
\aindex{Miranda, L. F.}     

%
%

\authormark{Maestro, Guerrero \& Miranda}

%
%



\begin{abstract}          
We present narrow-band images and long-slit echelle spectra of the
planetary nebula (PN) NGC\,6778.
The data show this PN as bipolar, with a very
prominent low-excitation equatorial toroid, high-excitation lobes and
two pairs of collimated outflows.
Morphologically, the pairs of outflows are different from each other;
one is linear and oriented along the bipolar axis, the other presents
an S-shape with changing orientations.
Besides the different morphology, both pairs of collimated outflows
present radial velocities increasing with distance from the central
star and share a common origin in bright knots at the tips
of the shell.
\end{abstract}

%
%

\section{Introduction}

NGC\,6778 is a planetary nebula (PN) of excitation class 5 whose
central star has Zanstra temperatures between 4.5$\times10^4$~K
(H) and 7.5$\times10^4$~K (He) (Preite-Martinez \& Pottasch 1983).
No molecular material (CO or H$_{2}$) has been detected in the
nebula (Huggins et al.\ 1996; Kastner et al.\  1996).
Electron density and electron temperature are $\simeq$1700~cm$^{-3}$
and $\simeq$10500~K, respectively (Preite-Martinez \& Pottasch 1983).
Its distance is uncertain, ranging from 1.0 to 3.7 kpc (see Acker et
al., 1992), with an averaged value of $\simeq$3~kpc.
Classified as a bipolar, filamentary PN by Peimbert \& Torres-Peimbert
(1983), the H$\alpha$+[N~{\sc ii}] and [O~{\sc iii}] images by Schwarz,
Corradi \& Melnick (1992) show an elliptical nebula with hints of
point-symmetric filaments.

We have obtained narrow-band images and high resolution long-slit
spectra of NGC\,6778 in order to analyze in detail its structure and
kinematics. In this paper, we present the preliminary results of our
investigation.

\section{Observations}

\begin{figure}[!ht]
\epsscale{0.7}
\plotone{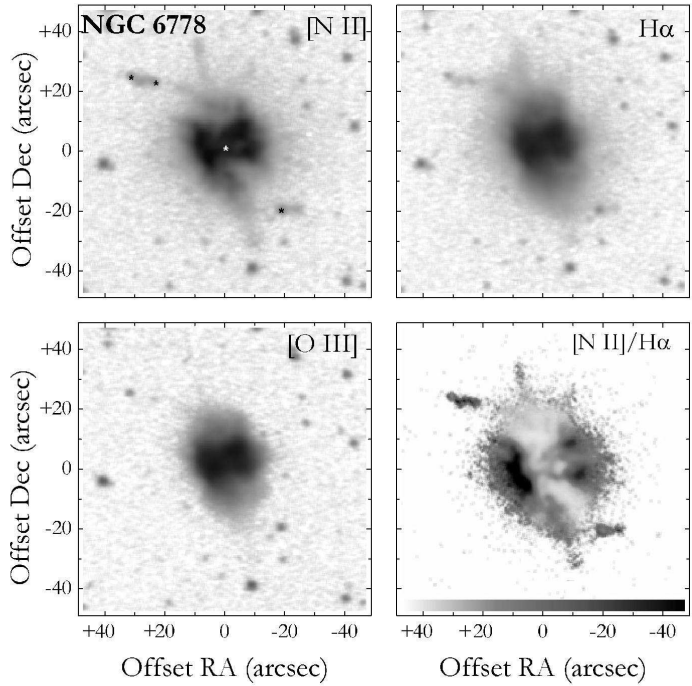}
\caption{Grey-scale [N~{\sc ii}], H$\alpha$ and [O~{\sc iii}] images and [N~{\sc ii}]/H$\alpha$
intesity ratio map of NGC\,
6778. Black dots in the [N~{\sc ii}] image indicate field stars, the white dot
marks the position of the central star, as deduced from the continuum
image (not shown here).}
\label{ima}
\end{figure}

Narrow-band images of NGC\,6778 were obtained on 2002 June 24 with the
1.5m telescope at Observatorio de Sierra Nevada (Granada, Spain).
The nebula was observed through narrow-band filters centered in the
H$\alpha$, [N~{\sc ii}] $\lambda$6583 and [O~{\sc iii}] $\lambda$5007
emission lines and in the continuum at $\lambda$6652 {\AA}. The detector was a
Wright CCD with 820$\times$1152 pixels and an image
scale of 0\farcs38~pixel$^{-1}$.
Exposure time ranged between 600~s and 1800~s. Seeing was $\simeq 1\farcs8$.

Long-slit echelle observations of NGC\,6778 were obtained on 2002 June
23 and 24 using the echelle spectrograph on the 4m telescope at the
Cerro Tololo Inter-American Observatory (CTIO).
The spectrograph was used in the single-order, long-slit mode covering
only the H$\alpha$ and [N~{\sc ii}] $\lambda\lambda$6548,6583 lines
with a slit length of $\sim$3$^\prime$.
The data were recorded with the SITE 2K $\#6$ CCD, resulting in a
spatial scale of $0\farcs26$~pixel$^{-1}$.
The instrumental FWHM  was $\sim$8~km~s$^{-1}$.
The slit was oriented at position angles (PAs) 15$^\circ$, 24$^\circ$,
47$^\circ$, and 100$^\circ$ to cover the different morphological features
of NGC\,6778.
Integration times were 1200~s at each slit position.

\section{Results}

\begin{figure}[!ht]
\epsscale{0.7}
\plotone{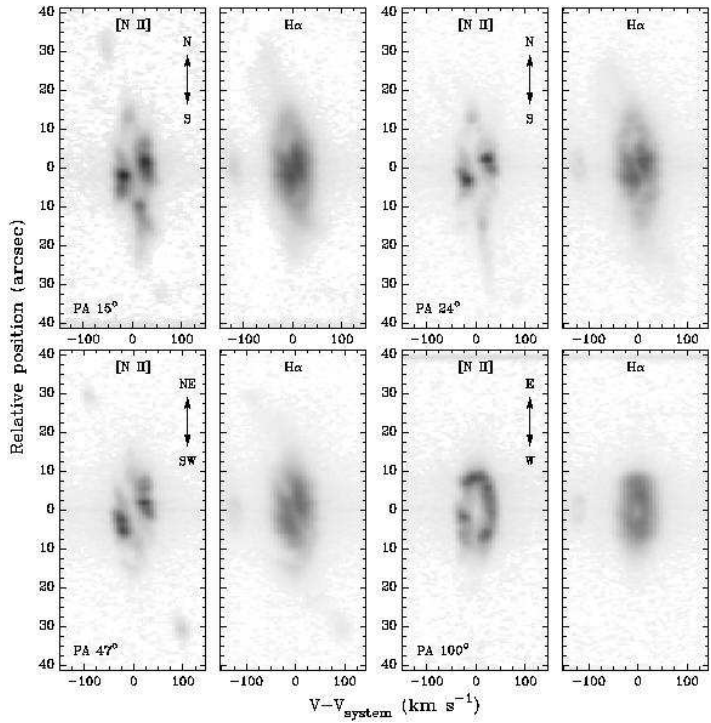} \caption{Grey-scale position-velocity maps of the
[N~{\sc ii}] and H$\alpha$ emission lines from NGC\,6778 observed at four different
PAs, shown at the bottom-left corner of the [N~{\sc ii}] pictures. The origin of the
vertical scale corresponds to the location of the central star, as implied from the
position of the stellar continuum peak.}
\label{spe}
\end{figure}

Figure \ref{ima} presents the [N~{\sc ii}], H$\alpha$, [O~{\sc iii}] images
and [N~{\sc ii}]/H$\alpha$ intensity ratio map of NGC\,6778.
The H$\alpha$ and [O~{\sc iii}] images show a bipolar PN of 20$''\times40''$
in size with its major axis oriented at PA$\sim$15$^\circ$, embedded in a
faint elliptical structure detected in H$\alpha$ with a size of
$\sim50''\times60''$.

The nebular emission is dominated by a bright and knotty equatorial region
of $\sim 16''\times10''$ in size.
In the light of [N~{\sc ii}], the emission is mostly confined to this
equatorial region.
Very interestingly, two pairs of jet-like features are detected in this
low-excitation line.
One pair has linear morphology oriented along the main axis of the bipolar
structure at PA$\sim$15$^\circ$ and can be traced up to $\simeq35''$ from
the central star.
The other pair presents an S-shape morphology between PA$\sim$15$^\circ$ and
$\sim$50$^\circ$ and can be traced up to $\simeq35''$ from the central star.

The [N~{\sc ii}]/H$\alpha$ intensity ratio map (Fig. \ref{ima}) shows that
low-ionization material is concentrated around the equatorial region and
in the jet-like features, while high-ionization material dominates the
bipolar lobes.


Figure \ref{spe} shows position-velocity maps of the four echellograms in the light
of [N~{\sc ii}] and H$\alpha$. The shapes of the emission lines in the position-velocity maps are
compatible with a bipolar shell.
The bright toroid observed in the direct images can be identified with the
brightest knots detected in the spectra.
If we assume circular cross-section for the toroid, its axis is tilted
by $\sim 15^\circ$ with respect to the plane of the sky, as determined
from its relative size along the major and minor axes of the nebula
observed in the spectra.

The jet-like features detected in the images are associated with high
velocity features in the long-slit spectra (Fig.~2), confirming that
they are collimated outflows.
The radial velocity of the outflow at PA $15^\circ$ increases linearly
with the distance to the central star from $\simeq$20~km\,s$^{-1}$ up
to $\simeq$60~km\,s$^{-1}$ with respect to the systemic velocity of NGC\,6778.
Assuming that this outflow shares the inclination of the bipolar shell,
its (maximum) deprojected velocity is $\simeq$230~km\,s$^{-1}$.
Similarly, the radial velocity of the S-shaped outflow increases with
the distance to the central star, with a maximum observed radial velocity
 of 100~km\,s$^{-1}$.

It is interesting to note that the collimated outflows seem
to originate from bright knots at the tips of the bipolar shell, as
it has already been observed in NGC\,6891 (Guerrero et al.\ 2000).
The likely connection between the collimated outflows and nebular features
suggests either that the shell has contributed to the collimation of the
outflows or that the outflows have interacted and shocked material in the
nebular shell.

\section{Conclusions}

Narrow-band images and long-slit echelle spectra indicate that NGC\,6778
is a bipolar PN with a very bright equatorial region.
Two pairs of collimated outflows have been detected in the object; one is
linear and oriented along the bipolar axis and the other one presents an
S-shape morphology.
The radial velocity in these outflows increases with the distance to the
central star, reaching values up to 100 km\,s$^{-1}$.
The outflows seem to arise from bright knots at the tips of the bipolar
shell suggesting that the shell has been involved in the collimation or
that the outflows have interacted with the nebular shell.

\vspace{0.5cm}

\noindent {\bf Acknowledgments.} VM and LFM acknowledge support from
MCyT grant (FEDER founds) AYA2002-00376 (Spain). This research was
partially based on data obtained at the Observatorio de Sierra Nevada
which is operated by the Consejo Superior de Investigaciones
Cient\'{\i}ficas through the Instituto de Astrof\'{\i}sica de
Andaluc\'{\i}a.

%
%
%
%



\begin{references}

\reference
Acker, A., \ et al.\ 1992, Strasbourg-ESO Catalogue of Galactic Planetary
Nebulae, ESO, Garching


\reference
Guerrero, M.\ A., Miranda, L.\ F., Manchado, A., \& V{\' a}zquez, R.\
2000, \mnras, 313, 1

\reference
Huggins, P.\ J., Bachiller, R., Cox, P., \& Forveille, T.\
1996, A\&A, 315, 284

\reference
Kastner, J.\ H., Weintraub, D., Gatley, I., Merrill, K.\ M., \& Probst,
R.\ G.\
1996, \apj, 462, 777

\reference
Peimbert, M., \& Torres-Peimbert, S.\
1983, IAU Symp.\ 103, 233

\reference
Preite-Martinez, A., \& Pottasch, S.\ R.,
1983, A\&A, 126, 31

\reference
Schwarz, H.\ E., Corradi, R.\ L.\ M., \& Melnick, J.\
1992, A\&AS, 96, 23



\end{references}
\end{document}